\documentclass[12pt]{iopart}
\usepackage{graphicx}
\usepackage{xcolor}

\begin{document}

\title[Multiphoton processes via conditional measurements in the two-field interaction]{Multiphoton processes via conditional measurements in the two-field interaction}

\author{J A Anaya-Contreras$^1$,  A Z\'u\~niga-Segundo$^1$,  A Perez-Leija$^2$, R de J Le\'on-Montiel$^3$ and H M Moya-Cessa$^4$ }

\address{$^1$ Instituto Polit\'ecnico Nacional, ESFM, Departamento de F\'isica. Edificio 9, Unidad Profesional “Adolfo L\'opez Mateos, 07738 CDMX., Mexico}
\address{$^2$Max Born Inst, Max Born Str. 2A, D-12489 Berlin, Germany}
\address{$^3$ Instituto de Ciencias Nucleares, Universidad Nacional Aut\'onoma de M\'exico, Apartado Postal 70-543, 04510 CDMX., Mexico}
\address{$^4$ Instituto Nacional de Astrof\'isica \'Optica y Electr\'onica, Luis Enrique Erro 1, Santa Mar\'ia Tonantzintla, Puebla, 72840 Mexico }
\ead{hmmc@inaoep.mx}

\begin{abstract}
In this contribution, we show that the use of conditional measurements in the resonant interaction of two quantized electromagnetic fields gives rise to nonclassical multiphoton processes. Furthermore, we demonstrate that this phenomenon may enable a robust integrated-optics protocol to engineer quantum states containing a high number of photons, thus making it a potentially appealing platform for exploring mesoscopic quantum phenomena.
\end{abstract}

%
%
%
%
%

\section{Introduction}

It is well known that conditional measurements  \cite{Kurizki1,Vogel,Garraway,Kurizki2,Foster}, i.e. projections of an entangled state onto a selected state once two subsystems have interacted, produce nonclassical states such as Schr\"odinger cat states \cite{Kurizki1} or more complex superpositions of coherent states \cite{Vogel,Garraway,Kurizki2,Dantsker} that may lead to generation of Fock states containing a high number of photons.\\

In the atom-field interaction, Gosh and Gerry \cite{Gosh} have shown that nonclassical states may be \textcolor{black}{engineered} by conditional measurements that produce gain or loss of several photons, even though the interaction is single-photon.  Luis \cite{Luis} called such gain of photons a paradoxical phenomenon and showed that a measurement leading to absorption of light produced an increase \textcolor{black}{in} the number of photons. Recently, Valverde and Baseia \cite{Baseia} studied the generation of nonclassical states \cite{Zarate} by such paradoxical processes in models of interpolating Hamiltonians, in particular by using nonlinear Jaynes-Cummings interactions. Anaya-Contreras {\it et al.} \cite{Anaya} have shown that it is possible to deposit several photons in the cavity by passing an atom through it and measuring its state as it exits the cavity. The fields produced showed ringing revivals of the atomic inversion  \cite{Optik,Satya} \textcolor{black}{commonly-observed} in the interaction of atoms with squeezed light \cite{Yuen,Caves,Loudon,Barnett2018}.\\

Interestingly, in recent years, the concept of conditional measurements has been successfully applied to the engineering of highly-correlated quantum optical states \cite{Carranza2012}. In this technique, the heralded subtraction of photons from two-mode squeezed vacuum states (TMSVS) gives rise to the generation of a family of correlated photon-subtracted TMSVS with a broad range of mean photon numbers and degrees of correlation \cite{kurochkin2014,omar2019}. Remarkably, this type of photon sources have shown to play an important role in the development of novel multiphoton integrated-optics protocols \cite{konrad1,konrad2}, as well as in \textcolor{black}{ quantum-enhanced spectroscopy \cite{svozilik2018}}, and metrology applications \cite{hofmann2006,birrittella2014,Barnett}.\\

In this contribution, we show that the generation of multiphoton quantum states via conditional measurements can be generalized to other systems such as the resonant interaction between two quantized fields. Given the simplicity of the model and its two-mode, beamsplitter-like nature, we anticipate its use in the implementation of multiphoton integrated-optics devices.\\

\section{Two-mode interaction with initial coherent states}

Consider the interaction Hamiltonian between two fields
\begin{equation}
	H=\lambda(a^{\dagger}b+b^{\dagger}a)\;,
\end{equation}
where $\lambda$ denotes the coupling strength between the two modes. \textcolor{black}{The interaction in Eq. (1) can be thought of as describing the interaction of the fields in a waveguide beamsplitter, where the coefficient $\lambda$ represents the hopping rate that results from the evanescent overlap between the normal modes supported by the waveguides \cite{Tailoring}. Note that the product $\lambda t$, with $t$ the propagation time (or length), defines the beamsplitter's output-port ratios \cite{konrad1}.} The evolution operator for the above Hamiltonian is then given by
\begin{equation}
	U(t)=\hbox{e}^{-iHt}=\hbox{e}^{-ia^{\dagger}b\tan \lambda t }\hbox{e}^{(b^{\dagger}b-a^{\dagger}a)\ln \cos\lambda t}\hbox{e}^{-ib^{\dagger}a\tan \lambda t }\;.
\end{equation}
If we consider the initial condition
\begin{equation} \label{initial}
	|\psi(0)\rangle = |\alpha\rangle_a|1\rangle_b\;,
\end{equation}
\textcolor{black}{{\it i.e.}, a coherent state in mode $a$ and the first excited (number state) state in mode $b$,} the solution reads \textcolor{black}{(for the sake of clearness, we show in Appendix A some details of the calculations)}
\begin{equation}
	\label{CHS2}
	|\psi(t) \rangle = U(t)|\psi(0)\rangle = \left(c b^{\dagger} \textcolor{black}{-} i s a^{\dagger}\right)|c\alpha\rangle_{a}|-is\alpha\rangle_{b}\;,
\end{equation}
with $c = \cos\lambda t $, and $s = \sin\lambda t$. If we detect $N$ photons in mode $b$ we collapse the wavefunction of mode $a$ to the state
\begin{equation}
	\label{CHS3}
	|\psi(t,N)\rangle_{a} = \frac{1}{\sqrt{N_{c}}}\left(\alpha_{0}(t,N)|c\alpha\rangle_{a} - i\alpha_{1}(t,N) a^{\dagger}|c\alpha\rangle_{a}\right)\;,
\end{equation}
with
\begin{equation} \label{CHS4}
	\alpha_{0}(t,N)=\hbox{e}^{-\frac{|\alpha|^{2}}{2}s^{2}}\frac{c(-i\alpha s)^{N-1}\sqrt{N}}{\sqrt{(N-1)!}}\;,
\end{equation}
\begin{equation} \label{CHS5}
	\alpha_{1}(t,N)=\hbox{e}^{-\frac{|\alpha|^{2}}{2}s^{2}}\frac{s(-i\alpha s)^{N}}{\sqrt{N!}}\;,
\end{equation}
and the normalization constant ${N_c}$ is given by
\begin{equation} \label{CHS6}
	N_{c} = |\alpha_{0}|^{2}+|\alpha_{1}|^{2}\left(1+|\alpha|^{2}c^{2}\right) +ic(\alpha\alpha_{0}\alpha^{*}_{1}-\alpha^{*}\alpha^{*}_{0}\alpha_{1}) \;.
\end{equation}

\subsection{Properties of the generated state}

The average number of photons  associated to the state (\ref{CHS3}) is given by
\begin{eqnarray} \label{CHS7}
	\bar{n}(t,N) &=& \frac{1}{N_{c}} \left(c^{2}\left|\alpha\right|^{2}\left|\alpha_{1}(t,N)\right|^{2}+\right.\nonumber\\
	&&+\left.\left|c\alpha\alpha_{0}(t,N)-i\left(1+c^{2}|\alpha|^{2}\right)\alpha_{1}(t,N)\right|^{2}\right)\;.
\end{eqnarray}

\begin{figure}[htbp]\label{Fig1}
	\centering
	{\includegraphics[scale=0.6]{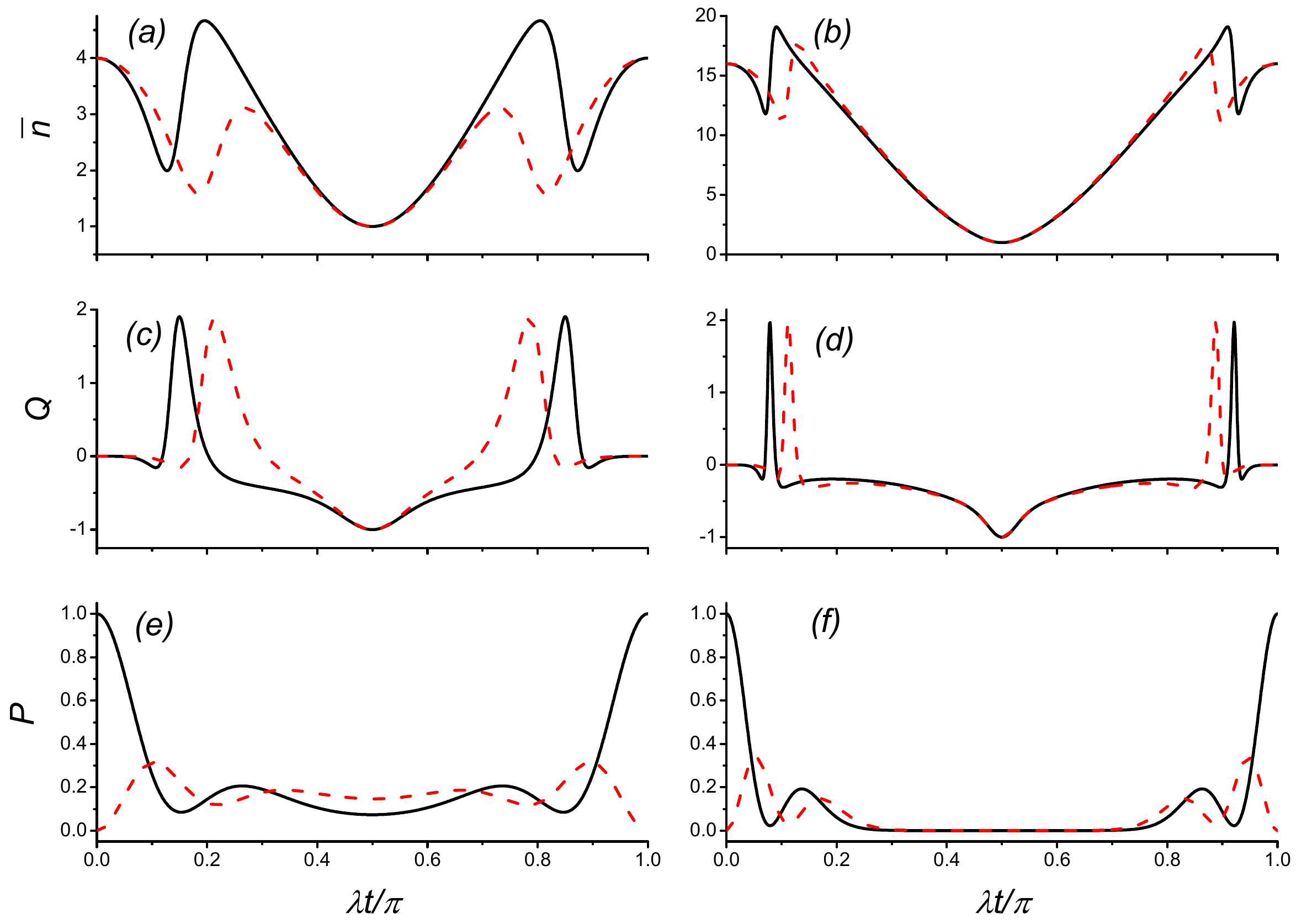}}
	\caption{(a) and (b): Plot
		of the average photon number for $N=1$ (black curve) and $N=2$ (red curve) and $\alpha=2$, and $\alpha=4$, respectively. (c) and (d): Plot
		of the \textcolor{black}{Mandel $Q$-parameter}  for $\alpha=2$ and $N=1$ (black curve) and $N=2$ (red curve) and $\alpha=4$, respectively. (e) and (f): Plot
		of the probability to detect one photon  (black curve) or two photons (red curve) in mode $b$ as a function of time, $\alpha=2$  and $\alpha=4$, respectively. } \label{Fig_1}
\end{figure}

The \textcolor{black}{Mandel $Q$-parameter} may be written in the form
\begin{eqnarray} \label{CHS8}
	Q_{M}(t,N)&=&\frac{c^{2}|\alpha|^{2}}{\bar{n}(t,N)N_{c}}\left(c^{2}|\alpha|^{2}\left|\alpha_{1}(t,N)\right|^{2}+\right.\nonumber\\ 
	&&\left.\left|c\alpha\alpha_{0}(t,N)-i\left(2+c^{2}|\alpha|^{2}\right)\alpha_{1}(t,N)\right|^{2}\right)-\bar{n}(t,N)\;.
\end{eqnarray}

The probability to detect  $N$ photons in the mode $b$ may be determined by the expression
\begin{equation} \label{CHS9}
	P(t,N)= |\alpha_{0}|^{2}+|\alpha_{1}|^{2}\left(1+|\alpha|^{2} c^{2}\right) +i c(\alpha\alpha_{0}\alpha^{*}_{1}-\alpha^{*}\alpha^{*}_{0}\alpha_{1}) \,,
\end{equation}
while the photon distribution for the state (\ref{CHS3})  is given by
\begin{equation} \label{CHS10}
	P_{n}(t,N)=\frac{\hbox{e}^{-|\alpha|^{2} c^{2}}( c |\alpha|)^{2n-2}}{n!N_{c}}\left| c\alpha\alpha_{0}(t,N)-in\alpha_{1}(t,N)\right|^{2} \;.
\end{equation}

\begin{figure}[h!]
	\centering
	{\includegraphics[scale=0.6]{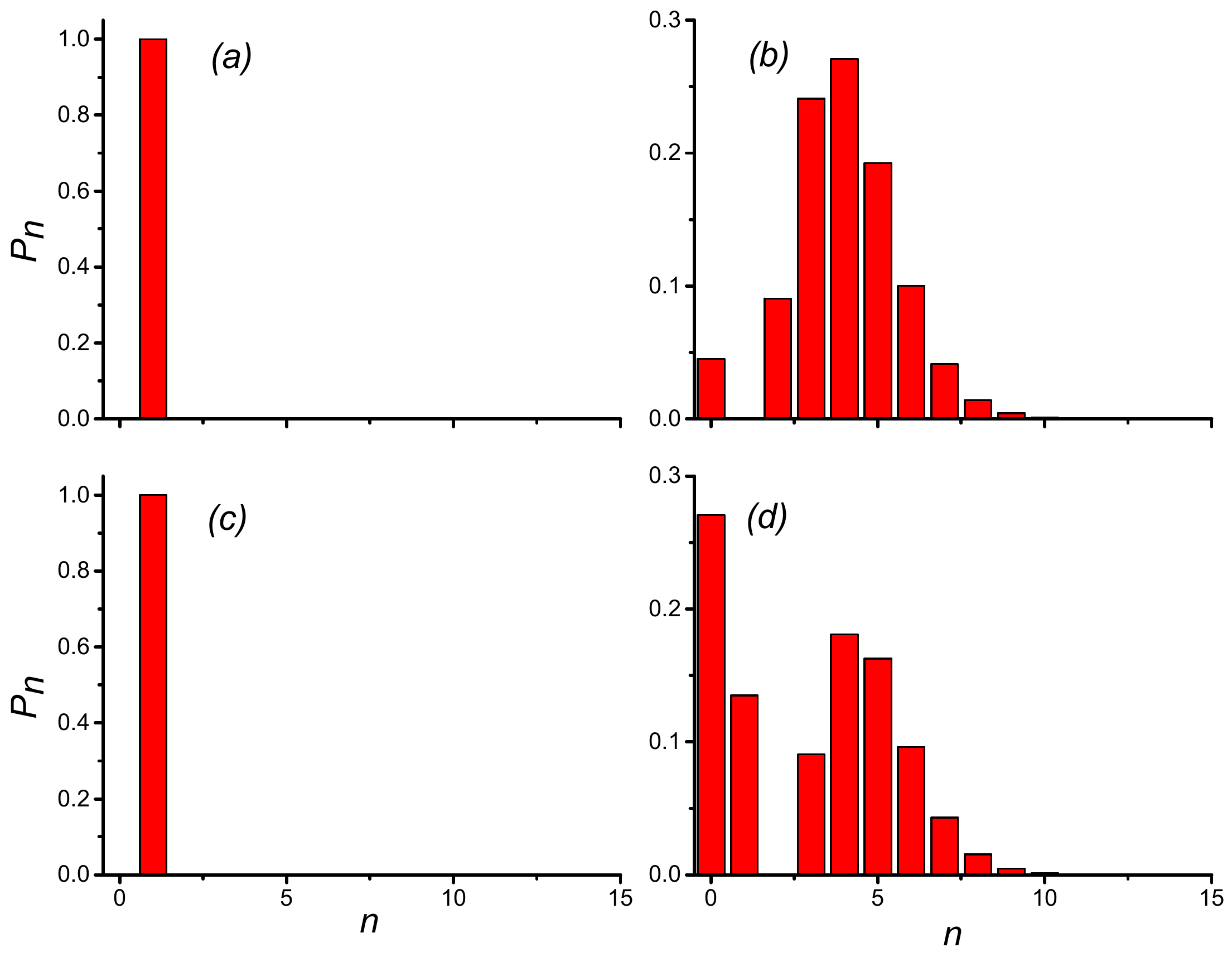}} 
	\caption{Plot
		of the photon distribution for  $|\alpha = 2\rangle$, 
		$N = 1$
		a) $\lambda t / \pi = 1/2$,
		b) $\lambda t /\pi = 1/4$,
		and  for 
		$N = 2$
		c) $\lambda t / \pi = 1/2$
		d) $\lambda t / \pi = 1/4$.} \label{Fig_2a}
\end{figure}

We plot in Fig. 1 the average number of photons, the \textcolor{black}{Mandel $Q$-parameter} and the probability of detecting one or two photons in mode $b$. The presented results show that multiphoton processes may be produced by the detection process, {\it i.e.,} although the interaction is a one-photon event, mode $a$ may \textcolor{black}{lose} more that one photon. \textcolor{black}{This may be clearly seen in Fig. 1(a,b) where, mode $a$ starts with four and sixteen photons, respectively, and loses several photons. In particular it may be seen at the first minimums the average number of photons of $\bar{n}\approx 2$ and $\bar{n}\approx 12$, respectively, when in mode $b$ a single photon  is measured (solid curve) and the effect is stronger when in mode $b$ two photons are measured (dashed line). The distribution of photons of the initial coherent state in mode $a$ allows that the field, although the process is a one-photon transition, loses more than one photon.  } Such kind of processes, of course, have low probability of occurring as shown in Figs.  1(e,f). The second row of Fig.  1 shows that number states may be generated, as the \textcolor{black}{Mandel $Q$-parameter} approaches the value minus one. \textcolor{black}{Figures 1(c,d) show that sub-Poissonian light is generated, as the \textcolor{black}{Mandel $Q$-parameter goes} below zero, for a wide range of interaction times. The photon distributions are plotted in Fig. 2 where the coherent nature of the field, although it keeps some of its original nature, differes from a Poissonian distribution.}

\section{\textcolor{black}{Thermal states}}

We show now the effect when a mixed state is considered in mode $a$ and a single photon in mode $b$, namely
\begin{equation} \label{THS1}
	\rho(0) = \sum_{m=0}^{\infty}\frac{P_{m}}{m!}\, b^{\dagger} a^{\dagger\,m} |0\rangle{_a}|0\rangle{_b}{_b}\langle0|{_a}\langle0| \,a^{m} b \;,
\end{equation}
which yields the \textcolor{black}{time-evolved} density matrix
\begin{eqnarray} \label{THS2}
	\rho(t) =  \sum_{m=0}^{\infty}\frac{P_{m}}{m!}\,\left( c b^{\dagger} - i s a^{\dagger}\right)&&\left(c a^{\dagger}-i s b^{\dagger} \right)^{m}\nonumber\\
	&& \times |0\rangle{_a}|0\rangle{_b}{_b}\langle0|{_a}\langle0| \,\left(c a+i s b\right)^{m} \left(c b + i s a\right)\;,
\end{eqnarray}
where, if we \textcolor{black}{consider that} $P_{m}=\frac{\bar{n}^{m}}{(1+\bar{n})^{m+1}}$, with $\bar{n}$ the average number of photons, describes an initial state given by a thermal distribution, the collapsed density matrix after $N$ photons are measured in mode $b$ is given by
\begin{eqnarray} \label{THS3}
	\rho_{a}(t,N) = \frac{1}{N_{T}}\sum_{m=0}^{\infty}\frac{P_{N+m-1}}{N+m}&&\left(\begin{array}{c}N+m\\N\end{array}\right)\nonumber\\
	&&\times c^{2m-2} s^{2N-2}\left(s^{2}m-c^{2}N\right)^{2}|m\rangle{_{a}}{_{a}}\langle m|\;,
\end{eqnarray}
with
\begin{equation} \label{THS4}
	N_{T} = \sum_{m=0}^{\infty}\frac{P_{N+m-1}}{N+m}\left(\begin{array}{c}N+m\\N\end{array}\right)c^{2m-2} s^{2N-2}\left(s^{2}m-c^{2}N\right)^{2}\;.
\end{equation}
\begin{figure}[h!]
	\centering
	{\includegraphics[scale=0.6]{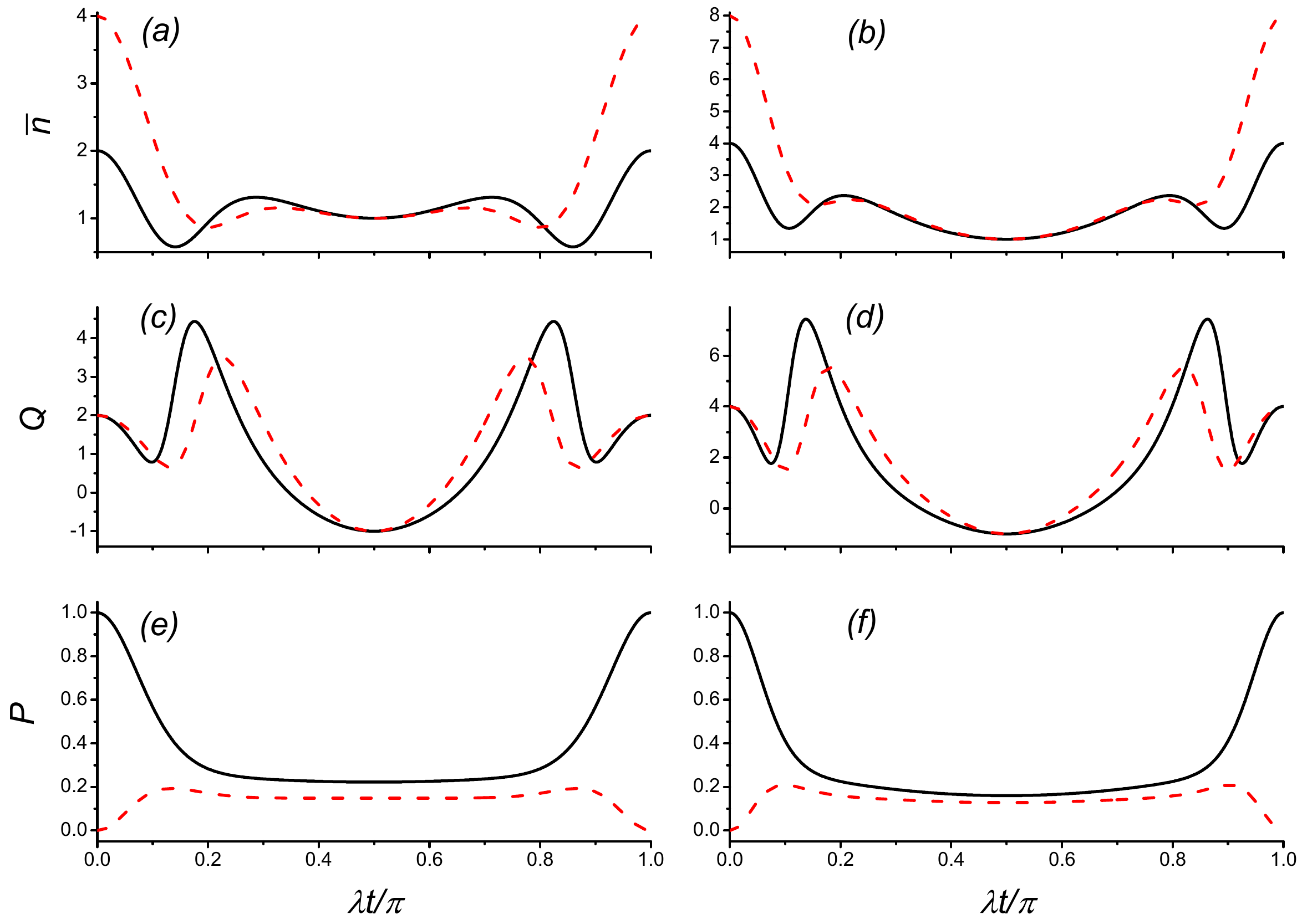}} 
	\caption{For a thermal distribution, (a) and (b): Plot
		of the average photon number for $N=1$ (black curve) and $N=2$ (red curve) and $\bar{n}=2$, and $\bar{n}=4$, respectively. (c) and (d): Plot
		of the \textcolor{black}{Mandel $Q$-parameter}  for $\bar{n}=2$ and $N=1$ (black curve) and $N=2$ (red curve) and $\bar{n}=4$, respectively. (e) and (f): Plot
		of the probability to detect one photon  (black curve) or two photons (red curve) in mode $b$ as a function of time, $\bar{n}=2$  and $\bar{n}=4$, respectively.} \label{Fig_2}
\end{figure}

\subsection{Properties of the generated state}

The average number of photons associated to the state (\ref{THS3}) is given by the expression
\begin{equation} \label{THS5}
	\bar{n}(t,N) =  \sum_{m=0}^{\infty}\frac{P_{N+m-1}}{N+m}\left(\begin{array}{c}N+m\\N\end{array}\right) c^{2m-2} s^{2N-2}\left( s^{2}m - c^{2}N\right)^{2}m\;.
\end{equation}

\begin{figure}[h!]
	\centering
	{\includegraphics[scale=0.6]{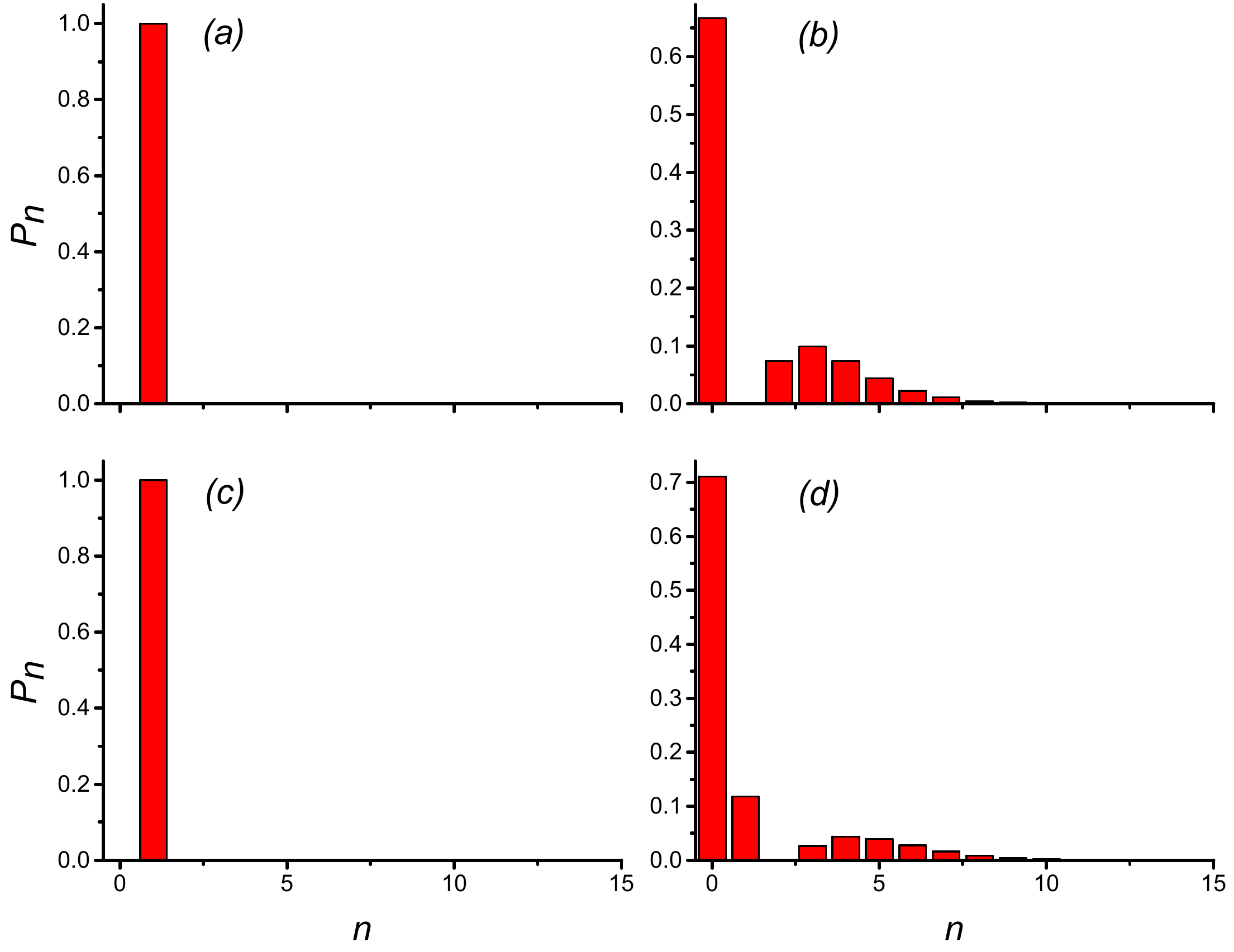}} 
	\caption{Plot
		of the photon distribution for  $\bar{n} = 2$, 
		$N = 1$
		a) $\lambda t / \pi = 1/2$,
		b) $\lambda t /\pi = 1/4$,
		and  for 
		$N = 2$
		c) $\lambda t / \pi = 1/2$,
		d) $\lambda t / \pi = 1/4$.} \label{Fig_2b}
\end{figure}
\noindent while the \textcolor{black}{Mandel $Q$-parameter} is
\begin{eqnarray} \label{THS6}
	Q_{M}(t,N) =-\bar{n}&&(t,N)+ \frac{1}{N_{T}\bar{n}(t,N)}\sum_{m=0}^{\infty}\frac{P_{N+m-1}}{N+m}
	\nonumber\\
	&&\times\left(\begin{array}{c}N+m\\N\end{array}\right)c^{2m-2}s^{2N-2}\left(s^{2}m-c^{2}N\right)^{2}m(m-1)\;.
\end{eqnarray}
and the probability of generating such state is given by
\begin{equation} \label{THS7}
	P(t,N) = \sum_{m=0}^{\infty}\frac{P_{N+m-1}}{N+m} \left(\begin{array}{c}N+m\\N\end{array}\right) c^{2m-2} s^{2N-2}\left(s^{2}m - c^{2}N\right)^{2}\;.
\end{equation}
\textcolor{black}{Finally, the photon} distribution may be easily found as
\begin{equation} \label{THS8}
	P_{n}(t,N)= \frac{1}{N_{T}}\frac{P_{N+n-1}}{N+n} \left(\begin{array}{c}N+n\\N\end{array}\right) c^{2n-2} s^{2N-2}\left(s^{2}n-c^{2}N\right)^{2} \;.
\end{equation}
\textcolor{black}{Figure 3, for an initial thermal field \textcolor{black}{in mode a}, shows the same behaviour as Fig. 1 for an initial coherent state, {\it i.e.}, the average number of photons [Fig. 3(a,b)] loses more than one photon for a wide range of interaction times. These events although the probability of occurrence is low, they still have some good probability (of around $0.25$) to occur [Fig. 3(e,f)]. As in the coherent state case, sub-Poissonian light is generated as seen in Fig. 3(c,d). \textcolor{black}{Similarly to the previous (coherent-state) case, the photon distributions show a reminiscent shape of the original, thermal distribution, but their behaviour departs from such distribution, as can be seen in Fig. 4(b,d).}}

\section{\textcolor{black}{Squeezed-vacuum states}}

We finally \textcolor{black}{consider} a squeezed vacuum state in mode $a$ and, again, a single photon in mode $b$ as initial state, {\i.e.},
\begin{equation} \label{SS1}
	|\psi(0)\rangle=\frac{1}{\sqrt{\cosh r}}\hbox{e}^{\frac{\tanh r}{2}a^{\dagger\,2}}|0\rangle{_a}|1\rangle{_b}\;.
\end{equation}
In this case, the \textcolor{black}{time-evolved} wavefunction gives
\begin{equation} \label{SS2}
	|\psi(t)\rangle = \frac{c b^{\dagger}-i s a^{\dagger}}{\sqrt{\cosh r}} \hbox{e}^{\frac{\tanh r}{2} c^{2} a^{\dagger\,2}}\hbox{e}^{-i c s \tanh r \, a^{\dagger} b^{\dagger}}\hbox{e}^{-\frac{\tanh r}{2} s^{2} b^{\dagger\,2}}|0\rangle{_a}|0\rangle{_b}\;,
\end{equation}
such that the \textcolor{black}{collapsed} wavefunction in mode $a$, after measuring $N$ photons in mode $b$, is given by
\begin{eqnarray} \label{SS3}
	|\psi(t,N)\rangle{_a}&&=\frac{c\sqrt{N}}{\sqrt{N_{sq}\cosh r}}\frac{1}{\sqrt{(N-1)!}}\times\nonumber\\
	&&\left(\sqrt{\frac{\tanh r}{2}} s \right)^{N-1}	H_{N-1}\left(-i\sqrt{\frac{\tanh r}{2}}c a^{\dagger}\right)\hbox{e}^{\frac{\tanh r}{2} c^{2} a^{\dagger\,2}}|0\rangle_{a}-\nonumber\\
	&&-\frac{i s}{\sqrt{N_{sq}\cosh r}}\frac{1}{\sqrt{N!}}\left(\sqrt{\frac{\tanh r}{2}} s \right)^{N}\times\nonumber\\
	&&\phantom{\frac{i s}{\sqrt{N_{sq}\cosh r}}\frac{1}{\sqrt{N!}}}H_{N}\left(-i\sqrt{\frac{\tanh r}{2}} c a^{\dagger}\right)\hbox{e}^{\frac{\tanh r}{2} c^{2} a^{\dagger\,2}} a^{\dagger}|0\rangle_{a}\;,
\end{eqnarray}
which may be written in a more compact form as
\begin{eqnarray} \label{SS4}
	|\psi(t,N)\rangle {_a} = \sqrt{\frac{2\tanh r N!}{N_{sq}\cosh r}}&&\left(\sqrt{\frac{\tanh r}{2}} s \right)^{N-1}\times\nonumber\\
	&&\sum_{m=0}^{N+1}Q_{m}(t,N)\, a^{\dagger\,m}\hbox{e}^{\frac{\tanh r}{2} c^{2} a^{\dagger\,2}}|0\rangle_{a}\;, 
\end{eqnarray}
with \footnote{$H_{n}(x)$ stand for the Hermite polynomials.}
\begin{eqnarray} \label{SS5}
	Q_{m}(t,N) = &&\frac{\left(-i\sqrt{2\tanh r} c \right)^{m-1}}{m!(N+1-m)!}\left(c^{2}(N-m)\times\phantom{\frac{s^{2}}{2}}\right.\nonumber\\
	&&\left.(N+1-m)H_{N-1-m}(0)+\frac{s^{2}}{2}mH_{N+1-m}(0)\right)\;,
\end{eqnarray}
\textcolor{black}{and}
\begin{eqnarray} \label{SS6}
	N_{sq} =\frac{2\tanh r N!}{\cosh r}\left(\frac{\tanh r}{2} s^{2}\right)^{N-1}\sum_{n=0}^{N+1}\sum_{m=0}^{N+1}Q^{*}_{n}(t,N)Q_{m}(t,N)\,I_{n,m}(t) \;,
\end{eqnarray}
\textcolor{black}{where (as we show in Appendix B)}
\begin{equation} \label{SS7}
	I_{n_{1},n_{2}}(t) = {_{a}}\langle0|e^{\frac{\tanh r}{2} c^{2} a^{2}} a^{n_{1}} a^{\dagger n_{2}}\hbox{e}^{\frac{\tanh r}{2} c^{2} a^{\dagger 2}}|0\rangle{_{a}}\;.
\end{equation}
We can now easily find the average number of photons associated to the state (\ref{SS4}), which is given by the expression
\begin{eqnarray} \label{SS8}
	\bar{n}(t,N) = \frac{2\tanh rN!}{N_{sq}\cosh r}&&\left(\frac{\tanh r}{2} s^{2}\right)^{N-1}\times\nonumber\\
	&&\sum_{n=0}^{N+1}\sum_{m=0}^{N+1}Q^{*}_{n}(t,N)Q_{m}(t,N)\,I_{n+1,m+1}(t)-1\;, 
\end{eqnarray}
while the \textcolor{black}{Mandel $Q$-parameter} is
\begin{equation} \label{SS9}
	Q_{M}(t,N)= \frac{{_a}\langle\psi(t,N)| a^{2} a^{\dagger\,2}|\psi(t,N)\rangle{_a}-2}{\bar{n}(t,N)}-\bar{n}(t,N)-4\;,
\end{equation}
with
\begin{eqnarray} \label{SS10}
	{_a}\langle\psi(t,N)| a^{2} a^{\dagger\,2}|\psi(t,N)\rangle{_a}=&& \frac{2\tanh rN!}{N_{sq}\cosh r}\left(\frac{\tanh r}{2} s^{2}\right)^{N-1}\times\nonumber\\
	&&\sum_{n=0}^{N+1}\sum_{m=0}^{N+1}Q^{*}_{n}(t,N)Q_{m}(t,N)\,I_{n+2,m+2}(t)\;.
\end{eqnarray}

\begin{figure}[h!]
	\centering
	{\includegraphics[scale=0.6]{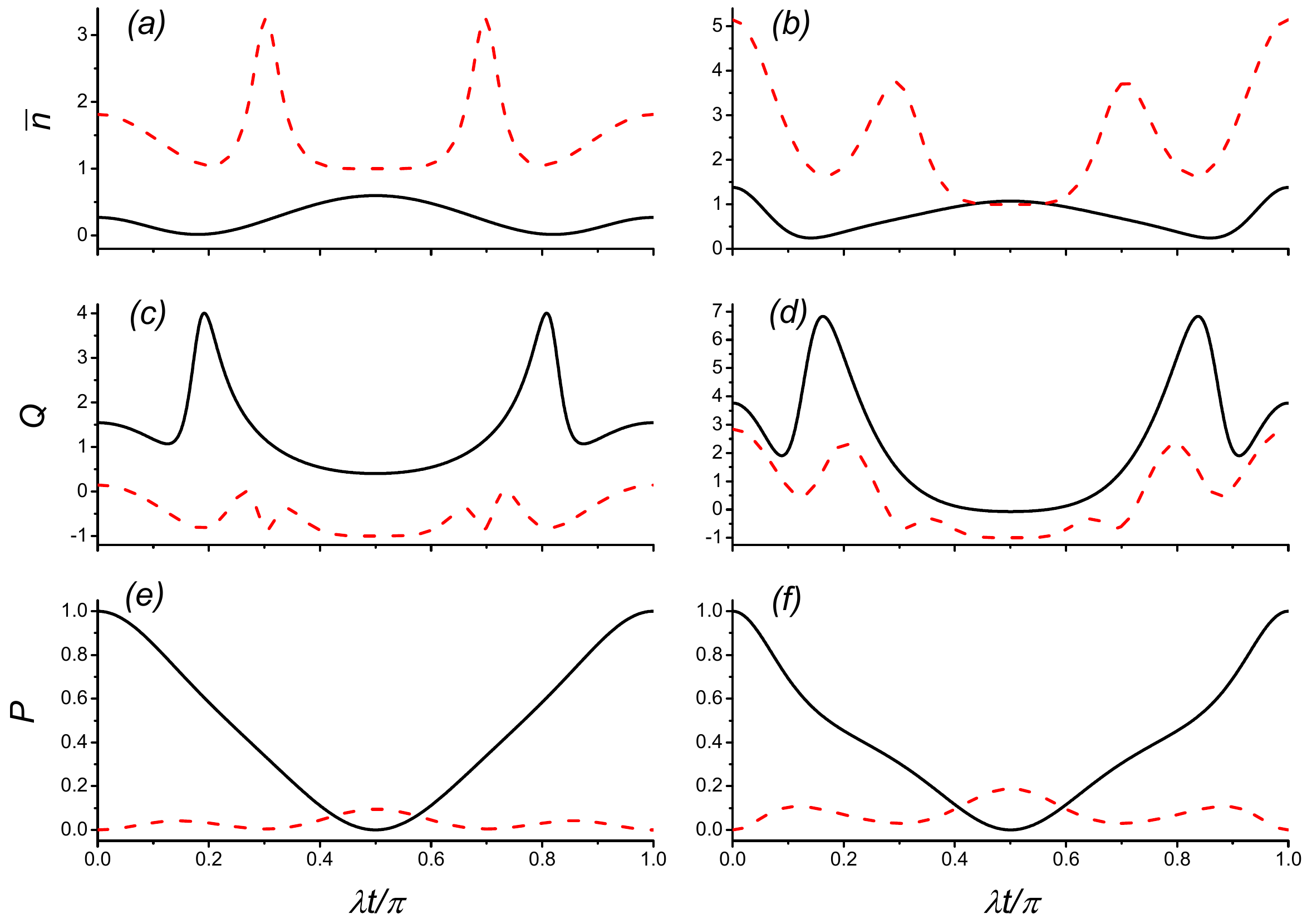}} 
	\caption{(a) and (b): Plot
		of the average photon number for $N=1$ (black curve) and $N=2$ (red curve) and $r=0.5$, and $r=1$, respectively. (c) and (d): Plot
		of the \textcolor{black}{Mandel $Q$-parameter}  for $r=0.5$ and $N=1$ (black curve) and $N=2$ (red curve) and $r=1$, respectively. (e) and (f): Plot
		of the probability to detect one photon  (black curve) or two photons (red curve) in mode $b$ as a function of time, $r=0.5$  and $r=1$, respectively.} \label{Fig_3}
\end{figure}

\noindent The probability to detect $N$ photons in mode $b$ is determined by
\begin{eqnarray} \label{SS11}
	P(t,N)=\frac{2\tanh rN!}{\cosh r}\left(\frac{\tanh r}{2} s^{2}\right)&&^{N-1}\times\nonumber\\
	&&\sum_{n=0}^{N+1}\sum_{m=0}^{N+1}Q^{*}_{n}(t,N)Q_{m}(t,N)\,I_{n,m}(t)\;.
\end{eqnarray}
with the photon distribution given as
\begin{eqnarray} \label{SS12}
	P_{n}(t,N)&=& \frac{2\tanh r N!}{N_{sq}\cosh r}\left(\frac{\tanh r}{2} s^{2}\right)^{N-1}\times\nonumber\\
	&&\sum_{l=0}^{\mathrm{min}(n,N+1)}\sum_{m=0}^{\mathrm{min}(n,N+1)} \frac{i^{l-m}n!H_{n-l}(0)H_{n-m}(0)}{(n-l)!(n-m)!} \times\nonumber\\
	&&\phantom{\sum_{l=0}^{\mathrm{min}(n,N+1)}} \times\left(\sqrt{\frac{\tanh r}{2}} c \right)^{2n-l-m}Q^{*}_{l}(t,N)Q_{m}(t,N) \;.
\end{eqnarray}

\begin{figure}[h!]
	\centering
	{\includegraphics[scale=0.6]{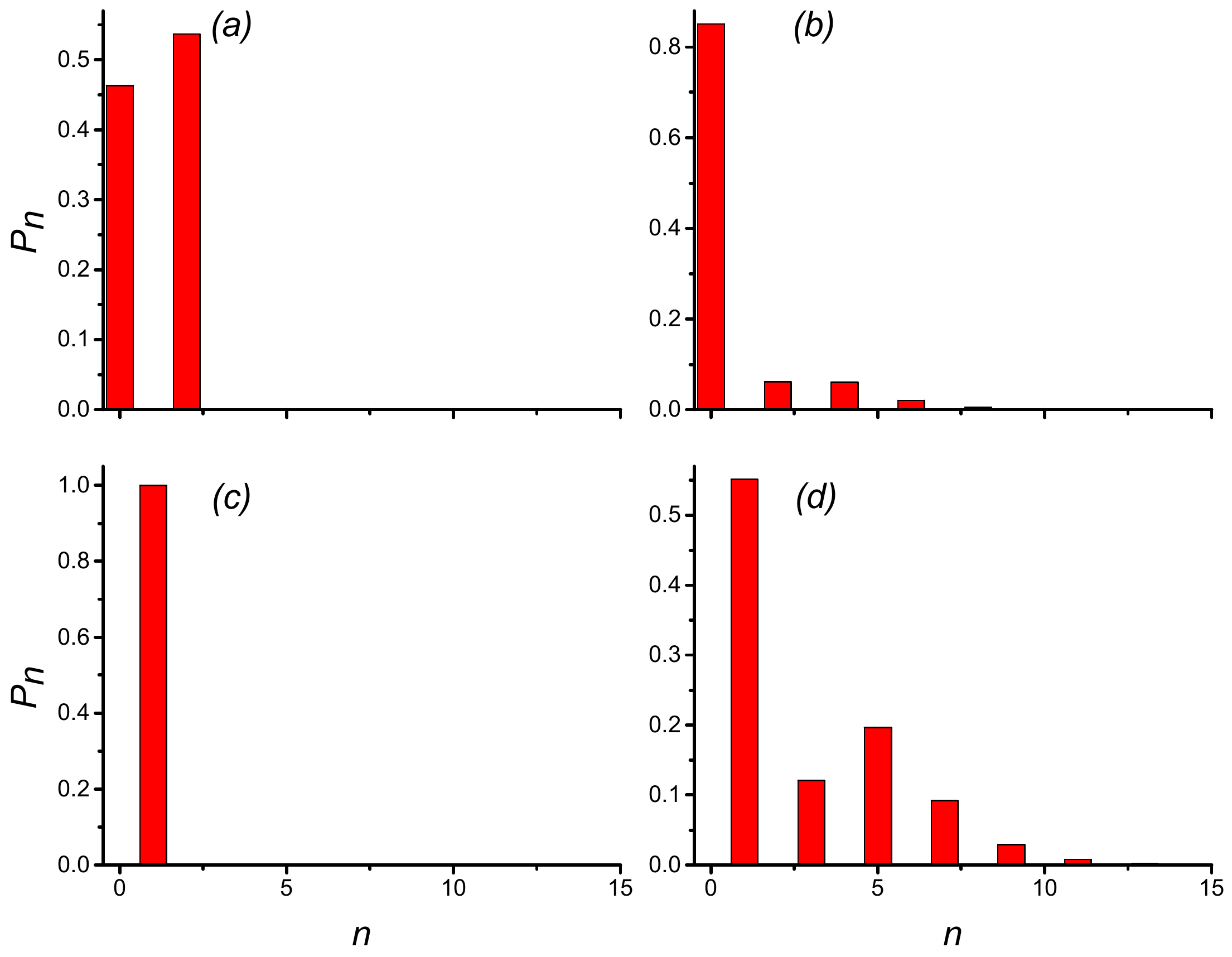}} 
	\caption{Plot of the photon distribution for  $|r = 1\rangle$, 
		$N = 1$
		a) $\lambda t / \pi = 1/2$,
		b) $\lambda t /\pi = 1/4$,
		and  for 
		$N = 2$
		c) $\lambda t / \pi = 1/2$
		d) $\lambda t / \pi = 1/4$.} \label{Fig_2c}
\end{figure}
Note that, as in the previous sections, the detection process in mode $b$ gives rise to the generation of multiphoton nonclassical states in port $a$. In particular, we can observe from Fig. 5 that by heralding the detection of $N=2$ photons in port $b$, we can generate sub-Poissonian states containing up to four photons in average. Although this process seems to have a rather small probability, one can always apply post-selection techniques to enhance the detection efficiency \cite{alan2019}.

\section{Conclusions}
We have shown that multiphoton nonclassical states may be generated by making use of conditional measurements in the resonant interaction of two fields.  \textcolor{black}{For the fields studied here, namely coherent, thermal and squeezed-vacuum states, we have shown that the conditional measurement in mode $b$ indeed generates nonclassical, sub-Poissonian light---witnessed by a Mandel $Q$-parameter below zero---in mode $a$.} Given the simplicity of the model and its beamsplitter-like nature, it can readily be used for exploring mesoscopic quantum phenomena in integrated optics devices.

\vfill\eject
   
\section*{Appendix A}
The {application} of the evolution operator on the annihilation and creation operator for both modes, $a$ and $b$ {gives}
\begin{equation}
    U(t)aU^{\dagger}(t)= ac(t)+i bs(t)\qquad 
    U(t)a^{\dagger}U^{\dagger}(t)= a^{\dagger}c(t)-i b^{\dagger}s(t),
\end{equation}
and 
\begin{equation}
    U(t)bU^{\dagger}(t)= bc(t) {+} i as(t)\qquad 
    U(t)b^{\dagger}U^{\dagger}(t)= b^{\dagger}c(t) -i a^{\dagger}s(t),
\end{equation}
respectively, with $c(t)=\cos\lambda t$ and $s(t)=\sin\lambda t$.
When we apply the evolution operator to the initial state (\ref{initial}){, i.e.,}
\begin{equation}
	|\psi(0)\rangle = b^{\dagger}e^{\alpha a^{\dagger}-\alpha^*a}|0\rangle_a|0\rangle_b\;,
\end{equation}
the {time-evolved} wavefunction becomes
\begin{equation}
	|\psi(t)\rangle =U(t)b^{\dagger} e^{\alpha a^{\dagger}-\alpha^*a}U^{\dagger}(t)U(t)|0\rangle_a|0\rangle_b\;.
\end{equation}
{Given that} $U(t)|0\rangle_a|0\rangle_b=|0\rangle_a|0\rangle_b$, we may write
\begin{equation}
	|\psi(t)\rangle =(cb^{\dagger}-isa^{\dagger}) e^{\alpha c a^{\dagger}-\alpha^*ca}e^{-i\alpha s b^{\dagger}-i\alpha^*sb}|0\rangle_a|0\rangle_b\;,
\end{equation}
such that we obtain
\begin{equation}
	|\psi(t)\rangle =(cb^{\dagger}-isa^{\dagger}) |c\alpha\rangle_a|-is\alpha\rangle_b\;.
\end{equation}
Now, by detecting $N$ photons in mode $b$ we obtain the wave function for mode $a$ as
\begin{equation}
_b\langle N	|\psi(t)\rangle =_b\langle N|(cb^{\dagger}-isa^{\dagger}) |c\alpha\rangle_a|-is\alpha\rangle_b\;,
\end{equation}
{which} may be rewritten as
\begin{equation}
	|\psi(t)\rangle_a =c\sqrt{N}_b\langle N-1| -is\alpha\rangle_b|c\alpha\rangle_a- isa^{\dagger}_b\langle N| -is\alpha\rangle_b|c\alpha\rangle_a,
\end{equation}
from where the Poissonian coefficients in equations  (\ref{CHS4}) y (\ref{CHS5}) are obtained.
\section*{Appendix B}
In this appendix we calculate the scalar product given in equation (\ref{SS7})
\begin{eqnarray} 
	\label{A1}
	I_{n_{1},n_{2}}(t) &=& \frac{1}{\pi} \int  \alpha^{n_{1}}\alpha^{*\,n_{2}}e^{-|\alpha|^{2}}e^{\frac{\tanh r}{2} c^{2}(\alpha^{2}+\alpha^{*\,2})}d^{2}\alpha \nonumber\\
	&=&\sum_{m_{1}=0}^{n_{1}}\sum_{m_{2}=0}^{n_{2}} 
	\left(\begin{array}{c}n_{1}\\m_{1}\end{array}\right)\left(\begin{array}{c}n_{2}\\m_{2}\end{array}\right)  i^{n_{1}-m_{2}-m_{1}}  \nonumber\\
	&\times&\frac{1}{\sqrt{\pi}}\int_{-\infty}^{\infty} e^{-(1- c^{2}\tanh r)\alpha^{2}_{x}}\alpha^{n_{2}+m_{1}-m_{2}}_{x}  d\alpha_{x} \nonumber\\
	&\times&\frac{1}{\sqrt{\pi}}\int_{-\infty}^{\infty}e^{-(1+ c^{2}\tanh r)\alpha^{2}_{y}}\alpha^{n_{1}+m_{2}-m_{1}}_{y} d\alpha_{y}\nonumber\\
	&=&\frac{1}{\sqrt{1-c^{4}\tanh^{2}r}} \sum_{m_{1}=0}^{n_{1}}\sum_{m_{2}=0}^{n_{2}} 
	\left(\begin{array}{c}n_{1}\\m_{1}\end{array}\right)\left(\begin{array}{c}n_{2}\\m_{2}\end{array}\right)
	i^{n_{1}-m_{2}-m_{1}}
	\nonumber\\
	&\times&
	\left(\frac{1}{\sqrt{1- c^{2}\tanh r}}\right)^{n_{2}+m_{1}-m_{2}}\left(\frac{1}{\sqrt{1+ c^{2}\tanh r}}\right)^{n_{1}+m_{2}-m_{1}}  \nonumber\\
	&\times&\frac{1}{\sqrt{\pi}}\int_{-\infty}^{\infty} x^{n_{2}+m_{1}-m_{2}}e^{-x^{2}}dx \frac{1}{\sqrt{\pi}} \int_{-\infty}^{\infty} y^{n_{1}+m_{2}-m_{1}}e^{-y^{2}}dy.
\end{eqnarray}
{By employing} the integral representation of Hermite polynomials\textcolor{black}{, i.e.,}
\begin{eqnarray}
	\frac{1}{\sqrt{\pi}}\int_{-\infty}^{\infty}t^{n}\hbox{e}^{-t^{2}}dt =\left(-\frac{i}{2}\right)^{n}H_{n}(0)
\end{eqnarray}
{equation (\ref{A1}) becomes}
\begin{eqnarray}
	\label{A2}
	&&I_{n_{1},n_{2}}(t)= \frac{1}{2^{n_{1}+n_{2}}}\frac{1}{\sqrt{1-c^{4}\tanh^{2}r}}\sum_{m_{1}=0}^{n_{1}}\sum_{m_{2}=0}^{n_{2}}\left(\begin{array}{c}n_{1}\\m_{1}\end{array}\right) \left(\begin{array}{c}n_{2}\\m_{2}\end{array}\right)\times \nonumber\\
	&\times &(-i)^{n_{2}+m_{1}+m_{2}}
	\left(\frac{1}{\sqrt{1-c^{2}\tanh r}}\right)^{n_{2}+m_{1}-m_{2}}\times \nonumber\\
	&\times&\left(\frac{1}{\sqrt{1+c^{2}\tanh r}}\right)^{n_{1}+m_{2}-m_{1}}  H_{n_{2}+m_{1}-m_{2}}(0)H_{n_{1}+m_{2}-m_{1}}(0)\,.
\end{eqnarray}

\end{document}